
  \input miniltx
  \def\Gin@driver{pdftex.def}
  \input color.sty
  \input graphicx.sty
  \resetatcatcode
%
%

%
%
%
%

\def\Serif{cmr}
\def\SerifBold{cmbx}
\def\SerifItalics{cmti}
\def\SerifSlanted{cmsl}
\def\SerifBoldItalics{cmbxti}
\def\SansSerif{cmss}
\def\SansSerifBold{cmssbx}
\def\SansSerifItalics{cmssi}
\def\SansSerifSlanted{cmssi}
\def\Math{cmmi}
\def\Symbols{cmsy}
\def\MathBold{cmmib}
\def\MoreSymbols{cmex}
\def\Typewriter{cmtt}
\def\Gothic{eufm}
\def\Double{msbm}
\def\Relazioni{msam}

= 			\Serif10 			at 5pt
= 		\SerifBold10 		at 5pt
= 	\SerifItalics10 	at 5pt
=		\SerifSlanted10 	at 5pt
=	\SerifBoldItalics10	at 5pt
= 		\SansSerif10 		at 5pt
=	\SansSerifBold10	at 5pt
=	\SansSerifItalics10	at 5pt
=	\SansSerifSlanted10	at 5pt
=				\Math10				at 5pt
=			\MathBold10			at 5pt
=			\Symbols10			at 5pt
=		\MoreSymbols10		at 5pt
=		\Typewriter10		at 5pt
=			\Gothic10			at 5pt
=			\Double10			at 5pt

= 			\Serif10 			at 7pt
= 		\SerifBold10 		at 7pt
= 	\SerifItalics10 	at 7pt
=	\SerifSlanted10 	at 7pt
=\SerifBoldItalics10	at 7pt
= 		\SansSerif10 		at 7pt
= 	\SansSerifBold10 	at 7pt
=\SansSerifItalics10	at 7pt
=\SansSerifSlanted10	at 7pt
=			\Math10				at 7pt
=		\MathBold10			at 7pt
=			\Symbols10			at 7pt
=		\MoreSymbols10		at 7pt
=		\Typewriter10		at 7pt
=			\Gothic10			at 7pt
=			\Double10			at 7pt

= 			\Serif10 			at 8pt
= 		\SerifBold10 		at 8pt
= 	\SerifItalics10 	at 8pt
=	\SerifSlanted10 	at 8pt
=\SerifBoldItalics10	at 8pt
= 		\SansSerif10 		at 8pt
= 	\SansSerifBold10 	at 8pt
=\SansSerifItalics10 at 8pt
=\SansSerifSlanted10 at 8pt
=			\Math10				at 8pt
=		\MathBold10			at 8pt
=			\Symbols10			at 8pt
=		\MoreSymbols10		at 8pt
=		\Typewriter10		at 8pt
=			\Gothic10			at 8pt
=			\Double10			at 8pt

= 			\Serif10 			at 10pt
= 		\SerifBold10 		at 10pt
= 		\SerifItalics10 	at 10pt
=		\SerifSlanted10 	at 10pt
=	\SerifBoldItalics10	at 10pt
= 		\SansSerif10 		at 10pt
= 	\SansSerifBold10 	at 10pt
= 	\SansSerifItalics10 at 10pt
= 	\SansSerifSlanted10 at 10pt
=				\Math10				at 10pt
=			\MathBold10			at 10pt
=			\Symbols10			at 10pt
=		\MoreSymbols10		at 10pt
=		\Typewriter10		at 10pt
=			\Gothic10			at 10pt
=			\Double10			at 10pt
=			\Relazioni10			at 10pt

= 				\Serif10 			at 12pt
= 			\SerifBold10 		at 12pt
= 		\SerifItalics10 	at 12pt
=		\SerifSlanted10 	at 12pt
=	\SerifBoldItalics10	at 12pt
= 			\SansSerif10 		at 12pt
= 		\SansSerifBold10 	at 12pt
= 	\SansSerifItalics10 at 12pt
= 	\SansSerifSlanted10 at 12pt
=				\Math10				at 12pt
=			\MathBold10			at 12pt
=			\Symbols10			at 12pt
=		\MoreSymbols10		at 12pt
=			\Typewriter10		at 12pt
=				\Gothic10			at 12pt
=				\Double10			at 12pt

= 			\Serif10 			at 14pt
= 		\SerifBold10 		at 14pt
= 	\SerifItalics10 	at 14pt
=		\SerifSlanted10 	at 14pt
=	\SerifBoldItalics10	at 14pt
= 		\SansSerif10 		at 14pt
= 	\SansSerifBold10 	at 14pt
= \SansSerifSlanted10 at 14pt
= \SansSerifItalics10 at 14pt
=				\Math10				at 14pt
=			\MathBold10			at 14pt
=			\Symbols10			at 14pt
=		\MoreSymbols10		at 14pt
=		\Typewriter10		at 14pt
=			\Gothic10			at 14pt
=			\Double10			at 14pt

\def\NormalStyle{\parindent=5pt\parskip=3pt\normalbaselineskip=14pt%
\def\nt{\tenSerif}%
\def\rm{\fam0\tenSerif}%
\textfont0=\tenSerif\scriptfont0=\sevenSerif\scriptscriptfont0=\fiveSerif
\textfont1=\tenMath\scriptfont1=\sevenMath\scriptscriptfont1=\fiveMath
\textfont2=\tenSymbols\scriptfont2=\sevenSymbols\scriptscriptfont2=\fiveSymbols
\textfont3=\tenMoreSymbols\scriptfont3=\sevenMoreSymbols\scriptscriptfont3=\fiveMoreSymbols
\textfont\itfam=\tenSerifItalics\def\it{\fam\itfam\tenSerifItalics}%
\textfont\slfam=\tenSerifSlanted\def\sl{\fam\slfam\tenSerifSlanted}%
\textfont\ttfam=\tenTypewriter\def\tt{\fam\ttfam\tenTypewriter}%
\textfont\bffam=\tenSerifBold%
\def\bf{\fam\bffam\tenSerifBold}\scriptfont\bffam=\sevenSerifBold\scriptscriptfont\bffam=\fiveSerifBold%
\def\cal{\tenSymbols}%
\def\greekbold{\tenMathBold}%
\def\gothic{\tenGothic}%
\def\Bbb{\tenDouble}%
\def\LieFont{\tenSerifItalics}%
\nt\normalbaselines\baselineskip=14pt%
}

\def\TitleStyle{\parindent=0pt\parskip=0pt\normalbaselineskip=15pt%
\def\nt{\fourteenSansSerifBold}%
\def\rm{\fam0\fourteenSansSerifBold}%
\textfont0=\fourteenSansSerifBold\scriptfont0=\tenSansSerifBold\scriptscriptfont0=\eightSansSerifBold
\textfont1=\fourteenMath\scriptfont1=\tenMath\scriptscriptfont1=\eightMath
\textfont2=\fourteenSymbols\scriptfont2=\tenSymbols\scriptscriptfont2=\eightSymbols
\textfont3=\fourteenMoreSymbols\scriptfont3=\tenMoreSymbols\scriptscriptfont3=\eightMoreSymbols
\textfont\itfam=\fourteenSansSerifItalics\def\it{\fam\itfam\fourteenSansSerifItalics}%
\textfont\slfam=\fourteenSansSerifSlanted\def\sl{\fam\slfam\fourteenSerifSansSlanted}%
\textfont\ttfam=\fourteenTypewriter\def\tt{\fam\ttfam\fourteenTypewriter}%
\textfont\bffam=\fourteenSansSerif%
\def\bf{\fam\bffam\fourteenSansSerif}\scriptfont\bffam=\tenSansSerif\scriptscriptfont\bffam=\eightSansSerif%
\def\cal{\fourteenSymbols}%
\def\greekbold{\fourteenMathBold}%
\def\gothic{\fourteenGothic}%
\def\Bbb{\fourteenDouble}%
\def\LieFont{\fourteenSerifItalics}%
\nt\normalbaselines\baselineskip=15pt%
}

\def\PartStyle{\parindent=0pt\parskip=0pt\normalbaselineskip=15pt%
\def\nt{\fourteenSansSerifBold}%
\def\rm{\fam0\fourteenSansSerifBold}%
\textfont0=\fourteenSansSerifBold\scriptfont0=\tenSansSerifBold\scriptscriptfont0=\eightSansSerifBold
\textfont1=\fourteenMath\scriptfont1=\tenMath\scriptscriptfont1=\eightMath
\textfont2=\fourteenSymbols\scriptfont2=\tenSymbols\scriptscriptfont2=\eightSymbols
\textfont3=\fourteenMoreSymbols\scriptfont3=\tenMoreSymbols\scriptscriptfont3=\eightMoreSymbols
\textfont\itfam=\fourteenSansSerifItalics\def\it{\fam\itfam\fourteenSansSerifItalics}%
\textfont\slfam=\fourteenSansSerifSlanted\def\sl{\fam\slfam\fourteenSerifSansSlanted}%
\textfont\ttfam=\fourteenTypewriter\def\tt{\fam\ttfam\fourteenTypewriter}%
\textfont\bffam=\fourteenSansSerif%
\def\bf{\fam\bffam\fourteenSansSerif}\scriptfont\bffam=\tenSansSerif\scriptscriptfont\bffam=\eightSansSerif%
\def\cal{\fourteenSymbols}%
\def\greekbold{\fourteenMathBold}%
\def\gothic{\fourteenGothic}%
\def\Bbb{\fourteenDouble}%
\def\LieFont{\fourteenSerifItalics}%
\nt\normalbaselines\baselineskip=15pt%
}

\def\ChapterStyle{\parindent=0pt\parskip=0pt\normalbaselineskip=15pt%
\def\nt{\fourteenSansSerifBold}%
\def\rm{\fam0\fourteenSansSerifBold}%
\textfont0=\fourteenSansSerifBold\scriptfont0=\tenSansSerifBold\scriptscriptfont0=\eightSansSerifBold
\textfont1=\fourteenMath\scriptfont1=\tenMath\scriptscriptfont1=\eightMath
\textfont2=\fourteenSymbols\scriptfont2=\tenSymbols\scriptscriptfont2=\eightSymbols
\textfont3=\fourteenMoreSymbols\scriptfont3=\tenMoreSymbols\scriptscriptfont3=\eightMoreSymbols
\textfont\itfam=\fourteenSansSerifItalics\def\it{\fam\itfam\fourteenSansSerifItalics}%
\textfont\slfam=\fourteenSansSerifSlanted\def\sl{\fam\slfam\fourteenSerifSansSlanted}%
\textfont\ttfam=\fourteenTypewriter\def\tt{\fam\ttfam\fourteenTypewriter}%
\textfont\bffam=\fourteenSansSerif%
\def\bf{\fam\bffam\fourteenSansSerif}\scriptfont\bffam=\tenSansSerif\scriptscriptfont\bffam=\eightSansSerif%
\def\cal{\fourteenSymbols}%
\def\greekbold{\fourteenMathBold}%
\def\gothic{\fourteenGothic}%
\def\Bbb{\fourteenDouble}%
\def\LieFont{\fourteenSerifItalics}%
\nt\normalbaselines\baselineskip=15pt%
}

\def\SectionStyle{\parindent=0pt\parskip=0pt\normalbaselineskip=13pt%
\def\nt{\twelveSansSerifBold}%
\def\rm{\fam0\twelveSansSerifBold}%
\textfont0=\twelveSansSerifBold\scriptfont0=\eightSansSerifBold\scriptscriptfont0=\eightSansSerifBold
\textfont1=\twelveMath\scriptfont1=\eightMath\scriptscriptfont1=\eightMath
\textfont2=\twelveSymbols\scriptfont2=\eightSymbols\scriptscriptfont2=\eightSymbols
\textfont3=\twelveMoreSymbols\scriptfont3=\eightMoreSymbols\scriptscriptfont3=\eightMoreSymbols
\textfont\itfam=\twelveSansSerifItalics\def\it{\fam\itfam\twelveSansSerifItalics}%
\textfont\slfam=\twelveSansSerifSlanted\def\sl{\fam\slfam\twelveSerifSansSlanted}%
\textfont\ttfam=\twelveTypewriter\def\tt{\fam\ttfam\twelveTypewriter}%
\textfont\bffam=\twelveSansSerif%
\def\bf{\fam\bffam\twelveSansSerif}\scriptfont\bffam=\eightSansSerif\scriptscriptfont\bffam=\eightSansSerif%
\def\cal{\twelveSymbols}%
\def\bg{\twelveMathBold}%
\def\gothic{\twelveGothic}%
\def\Bbb{\twelveDouble}%
\def\LieFont{\twelveSerifItalics}%
\nt\normalbaselines\baselineskip=13pt%
}

\def\SubSectionStyle{\parindent=0pt\parskip=0pt\normalbaselineskip=13pt%
\def\nt{\twelveSansSerifItalics}%
\def\rm{\fam0\twelveSansSerifItalics}%
\textfont0=\twelveSansSerifItalics\scriptfont0=\eightSansSerifItalics\scriptscriptfont0=\eightSansSerifItalics%
\textfont1=\twelveMath\scriptfont1=\eightMath\scriptscriptfont1=\eightMath%
\textfont2=\twelveSymbols\scriptfont2=\eightSymbols\scriptscriptfont2=\eightSymbols%
\textfont3=\twelveMoreSymbols\scriptfont3=\eightMoreSymbols\scriptscriptfont3=\eightMoreSymbols%
\textfont\itfam=\twelveSansSerif\def\it{\fam\itfam\twelveSansSerif}%
\textfont\slfam=\twelveSansSerifSlanted\def\sl{\fam\slfam\twelveSerifSansSlanted}%
\textfont\ttfam=\twelveTypewriter\def\tt{\fam\ttfam\twelveTypewriter}%
\textfont\bffam=\twelveSansSerifBold%
\def\bf{\fam\bffam\twelveSansSerifBold}\scriptfont\bffam=\eightSansSerifBold\scriptscriptfont\bffam=\eightSansSerifBold%
\def\cal{\twelveSymbols}%
\def\greekbold{\twelveMathBold}%
\def\gothic{\twelveGothic}%
\def\Bbb{\twelveDouble}%
\def\LieFont{\twelveSerifItalics}%
\nt\normalbaselines\baselineskip=13pt%
}

\def\AuthorStyle{\parindent=0pt\parskip=0pt\normalbaselineskip=14pt%
\def\nt{\tenSerif}%
\def\rm{\fam0\tenSerif}%
\textfont0=\tenSerif\scriptfont0=\sevenSerif\scriptscriptfont0=\fiveSerif
\textfont1=\tenMath\scriptfont1=\sevenMath\scriptscriptfont1=\fiveMath
\textfont2=\tenSymbols\scriptfont2=\sevenSymbols\scriptscriptfont2=\fiveSymbols
\textfont3=\tenMoreSymbols\scriptfont3=\sevenMoreSymbols\scriptscriptfont3=\fiveMoreSymbols
\textfont\itfam=\tenSerifItalics\def\it{\fam\itfam\tenSerifItalics}%
\textfont\slfam=\tenSerifSlanted\def\sl{\fam\slfam\tenSerifSlanted}%
\textfont\ttfam=\tenTypewriter\def\tt{\fam\ttfam\tenTypewriter}%
\textfont\bffam=\tenSerifBold%
\def\bf{\fam\bffam\tenSerifBold}\scriptfont\bffam=\sevenSerifBold\scriptscriptfont\bffam=\fiveSerifBold%
\def\cal{\tenSymbols}%
\def\greekbold{\tenMathBold}%
\def\gothic{\tenGothic}%
\def\Bbb{\tenDouble}%
\def\LieFont{\tenSerifItalics}%
\nt\normalbaselines\baselineskip=14pt%
}


\def\AbstractStyle{\parindent=0pt\parskip=0pt\normalbaselineskip=12pt%
\def\nt{\eightSerif}%
\def\rm{\fam0\eightSerif}%
\textfont0=\eightSerif\scriptfont0=\sevenSerif\scriptscriptfont0=\fiveSerif
\textfont1=\eightMath\scriptfont1=\sevenMath\scriptscriptfont1=\fiveMath
\textfont2=\eightSymbols\scriptfont2=\sevenSymbols\scriptscriptfont2=\fiveSymbols
\textfont3=\eightMoreSymbols\scriptfont3=\sevenMoreSymbols\scriptscriptfont3=\fiveMoreSymbols
\textfont\itfam=\eightSerifItalics\def\it{\fam\itfam\eightSerifItalics}%
\textfont\slfam=\eightSerifSlanted\def\sl{\fam\slfam\eightSerifSlanted}%
\textfont\ttfam=\eightTypewriter\def\tt{\fam\ttfam\eightTypewriter}%
\textfont\bffam=\eightSerifBold%
\def\bf{\fam\bffam\eightSerifBold}\scriptfont\bffam=\sevenSerifBold\scriptscriptfont\bffam=\fiveSerifBold%
\def\cal{\eightSymbols}%
\def\greekbold{\eightMathBold}%
\def\gothic{\eightGothic}%
\def\Bbb{\eightDouble}%
\def\LieFont{\eightSerifItalics}%
\nt\normalbaselines\baselineskip=12pt%
}

\def\RefsStyle{\parindent=0pt\parskip=0pt%
\def\nt{\eightSerif}%
\def\rm{\fam0\eightSerif}%
\textfont0=\eightSerif\scriptfont0=\sevenSerif\scriptscriptfont0=\fiveSerif
\textfont1=\eightMath\scriptfont1=\sevenMath\scriptscriptfont1=\fiveMath
\textfont2=\eightSymbols\scriptfont2=\sevenSymbols\scriptscriptfont2=\fiveSymbols
\textfont3=\eightMoreSymbols\scriptfont3=\sevenMoreSymbols\scriptscriptfont3=\fiveMoreSymbols
\textfont\itfam=\eightSerifItalics\def\it{\fam\itfam\eightSerifItalics}%
\textfont\slfam=\eightSerifSlanted\def\sl{\fam\slfam\eightSerifSlanted}%
\textfont\ttfam=\eightTypewriter\def\tt{\fam\ttfam\eightTypewriter}%
\textfont\bffam=\eightSerifBold%
\def\bf{\fam\bffam\eightSerifBold}\scriptfont\bffam=\sevenSerifBold\scriptscriptfont\bffam=\fiveSerifBold%
\def\cal{\eightSymbols}%
\def\greekbold{\eightMathBold}%
\def\gothic{\eightGothic}%
\def\Bbb{\eightDouble}%
\def\LieFont{\eightSerifItalics}%
\nt\normalbaselines\baselineskip=10pt%
}

\def\ClaimStyle{\parindent=5pt\parskip=3pt\normalbaselineskip=14pt%
\def\nt{\tenSerifSlanted}%
\def\rm{\fam0\tenSerifSlanted}%
\textfont0=\tenSerifSlanted\scriptfont0=\sevenSerifSlanted\scriptscriptfont0=\fiveSerifSlanted
\textfont1=\tenMath\scriptfont1=\sevenMath\scriptscriptfont1=\fiveMath
\textfont2=\tenSymbols\scriptfont2=\sevenSymbols\scriptscriptfont2=\fiveSymbols
\textfont3=\tenMoreSymbols\scriptfont3=\sevenMoreSymbols\scriptscriptfont3=\fiveMoreSymbols
\textfont\itfam=\tenSerifItalics\def\it{\fam\itfam\tenSerifItalics}%
\textfont\slfam=\tenSerif\def\sl{\fam\slfam\tenSerif}%
\textfont\ttfam=\tenTypewriter\def\tt{\fam\ttfam\tenTypewriter}%
\textfont\bffam=\tenSerifBold%
\def\bf{\fam\bffam\tenSerifBold}\scriptfont\bffam=\sevenSerifBold\scriptscriptfont\bffam=\fiveSerifBold%
\def\cal{\tenSymbols}%
\def\greekbold{\tenMathBold}%
\def\gothic{\tenGothic}%
\def\Bbb{\tenDouble}%
\def\LieFont{\tenSerifItalics}%
\nt\normalbaselines\baselineskip=14pt%
}


%
%


\def\ModeYes{yes}
\def\ModeNo{no}

\def\ModeUndef{undefined}


\def\nx{\noexpand}
\def\ni{\noindent}
\def\newpage{\vfill\eject}

\def\ss{\vskip 5pt}
\def\ms{\vskip 10pt}
\def\bs{\vskip 20pt}

 \def\,{\mskip\thinmuskip}
 \def\!{\mskip-\thinmuskip}
 \def\>{\mskip\medmuskip}
 \def\;{\mskip\thickmuskip}

%
% ... 
%

\def\refsModePost{post}
\def\refsModeAuto{auto}

\def\dbRefsSatusModeOk{ok}
\def\dbRefsSatusModeError{error}
\def\dbRefsSatusModeWarning{warning}


\newcount\BNUM
\BNUM=0

\def\refs{}

\def\SetModePost{\xdef\refsMode{\refsModePost}}			
\SetModePost

\def\dbRefsStatusOk{%
	\xdef\dbRefsStatus{\dbRefsSatusModeOk}%
	\xdef\dbRefsError{\ModeNo}%
	\xdef\dbRefsWarning{\ModeNo}%
	\xdef\dbRefsInfo{\ModeNo}%
}

\def\dbRefs{%
}

\def\dbRefsGet#1{%
	\xdef\found{N}\xdef\ikey{#1}\dbRefsStatusOk%
	\xdef\key{\ModeUndef}\xdef\tag{\ModeUndef}\xdef\tail{\ModeUndef}%
	\dbRefs%
}

\def\NextRefsTag{%
	\global\advance\BNUM by 1%
}
\def\ShowTag#1{{\bf [#1]}}

\def\dbRefsInsert#1#2{%
\dbRefsGet{#1}%
\if\found Y %
   \xdef\dbRefsStatus{\dbRefsSatusModeWarning}%
   \xdef\dbRefsWarning{record is already there}%
   \xdef\dbRefsInfo{record not inserted}%
\else%
   \toks2=\expandafter{\dbRefs}%
   \ifx\refsMode\refsModeAuto \NextRefsTag
    \xdef\dbRefs{%
   	\the\toks2 \nx\xdef\nx\dbx{#1}%
	\nx\ifx\nx\ikey %
		\nx\dbx\nx\xdef\nx\found{Y}%
		\nx\xdef\nx\key{#1}%
		\nx\xdef\nx\tag{\the\BNUM}%
		\nx\xdef\nx\tail{#2}%
	\nx\fi}%
	\global\xdef\refs{\refs \ss\ni[\the\BNUM]\ #2\par}
   \fi%
   \ifx\refsMode\refsModePost 
    \xdef\dbRefs{%
   	\the\toks2 \nx\xdef\nx\dbx{#1}%
	\nx\ifx\nx\ikey %
		\nx\dbx\nx\xdef\nx\found{Y}%
		\nx\xdef\nx\key{#1}%
		\nx\xdef\nx\tag{\ModeUndef}%
		\nx\xdef\nx\tail{#2}%
	\nx\fi}%
   \fi%
\fi%
}

\def\dbRefsEdit#1#2#3{\dbRefsGet{#1}%
\if\found N 
   \xdef\dbRefsStatus{\dbRefsSatusModeError}%
   \xdef\dbRefsError{record is not there}%
   \xdef\dbRefsInfo{record not edited}%
\else%
   \toks2=\expandafter{\dbRefs}%
   \xdef\dbRefs{\the\toks2%
   \nx\xdef\nx\dbx{#1}%
   \nx\ifx\nx\ikey\nx\dbx %
	\nx\xdef\nx\found{Y}%
	\nx\xdef\nx\key{#1}%
	\nx\xdef\nx\tag{#2}%
	\nx\xdef\nx\tail{#3}%
   \nx\fi}%
\fi%
}

\def\bib#1#2{\RefsStyle\dbRefsInsert{#1}{#2}%
	\ifx\dbRefsStatus\dbRefsSatusModeWarning %
		\message{^^J}%
		\message{WARNING: Reference [#1] is doubled.^^J}%
	\fi%
}

\def\ref#1{\dbRefsGet{#1}%
\ifx\found N %
  \message{^^J}%
  \message{ERROR: Reference [#1] unknown.^^J}%
  \ShowTag{??}%
\else%
	\ifx\tag\ModeUndef \NextRefsTag%
		\dbRefsEdit{#1}{\the\BNUM}{\tail}%
		\dbRefsGet{#1}%
		\global\xdef\refs{\refs \ss\ni [\tag]\ \tail\par}
	\fi
	\ShowTag{\tag}%
\fi%
}

\def\ShowBiblio{\ms\Ensure{\SectionEnsure}%
{\SectionStyle\ni References}%
{\RefsStyle\refs}%
}

\newcount\CHANGES
\CHANGES=0
\def\AuxFile{7}
\def\PreventDoubleOn{\xdef\PreventDoubleLabel{\ModeYes}}

\PreventDoubleOn

\def\StoreLabel#1#2{\xdef\itag{#2}
 \ifx\PreModeStatus\ModeNo %
   \message{^^J}%
   \errmessage{You can't use Check without starting with OpenPreMode (and finishing with ClosePreMode)^^J}%
 \else%
   \immediate\write\AuxFile{\nx\dbLabelPreInsert{#1}{\itag}}%
   \dbLabelGet{#1}%
   \ifx\itag\tag %
   \else%
	\global\advance\CHANGES by 1%
 	\xdef\itag{(?.??)}%
    \fi%
   \fi%
}

\def\PreModeStatus{\ModeNo}

\def\ClosePreMode{\immediate\closeout\AuxFile%
  \ifnum\CHANGES=0%
	\message{^^J}%
	\message{**********************************^^J}%
	\message{**  NO CHANGES TO THE AuxFile  **^^J}%
	\message{**********************************^^J}%
 \else%
	\message{^^J}%
	\message{**************************************************^^J}%
	\message{**  PLAEASE TYPESET IT AGAIN (\the\CHANGES)  **^^J}%
    \errmessage{**************************************************^^ J}%
  \fi%
  \edef\PreModeStatus{\ModeNo}%
}

\def\dbLabelSatusModeOk{ok}

\def\dbLabelSatusModeWarning{warning}

\def\dbLabelStatusOk{%
	\xdef\dbLabelStatus{\dbLabelSatusModeOk}%
	\xdef\dbLabelError{\ModeNo}%
	\xdef\dbLabelWarning{\ModeNo}%
	\xdef\dbLabelInfo{\ModeNo}%
}

\def\dbLabel{%
}

\def\dbLabelGet#1{%
	\xdef\found{N}\xdef\ikey{#1}\dbLabelStatusOk%
	\xdef\key{\ModeUndef}\xdef\tag{\ModeUndef}\xdef\pre{\ModeUndef}%
	\dbLabel%
}

\def\ShowLabel#1{%
 \dbLabelGet{#1}%
 \ifx\tag \ModeUndef %
 	\global\advance\CHANGES by 1%
 	(?.??)%
 \else%
 	\tag%
 \fi%
}

\def\dbLabelPreInsert#1#2{\dbLabelGet{#1}%
\if\found Y %
  \xdef\dbLabelStatus{\dbLabelSatusModeWarning}%
   \xdef\dbLabelWarning{Label is already there}%
   \xdef\dbLabelInfo{Label not inserted}%
   \message{^^J}%
   \errmessage{Double pre definition of label [#1]^^J}%
\else%
   \toks2=\expandafter{\dbLabel}%
    \xdef\dbLabel{%
   	\the\toks2 \nx\xdef\nx\dbx{#1}%
	\nx\ifx\nx\ikey %
		\nx\dbx\nx\xdef\nx\found{Y}%
		\nx\xdef\nx\key{#1}%
		\nx\xdef\nx\tag{#2}%
		\nx\xdef\nx\pre{\ModeYes}%
	\nx\fi}%
\fi%
}

\def\dbLabelInsert#1#2{\dbLabelGet{#1}%
\xdef\itag{#2}%
\dbLabelGet{#1}%
\if\found Y %
	\ifx\tag\itag %
	\else%
	   \ifx\PreventDoubleLabel\ModeYes %
		\message{^^J}%
		\errmessage{Double definition of label [#1]^^J}%
	   \else%
		\message{^^J}%
		\message{Double definition of label [#1]^^J}%
	   \fi%
	\fi%
   \xdef\dbLabelStatus{\dbLabelSatusModeWarning}%
   \xdef\dbLabelWarning{Label is already there}%
   \xdef\dbLabelInfo{Label not inserted}%
\else%
   \toks2=\expandafter{\dbLabel}%
    \xdef\dbLabel{%
   	\the\toks2 \nx\xdef\nx\dbx{#1}%
	\nx\ifx\nx\ikey %
		\nx\dbx\nx\xdef\nx\found{Y}%
		\nx\xdef\nx\key{#1}%
		\nx\xdef\nx\tag{#2}%
		\nx\xdef\nx\pre{\ModeNo}%
	\nx\fi}%
\fi%
}


\newcount\PART
\newcount\CHAPTER
\newcount\SECTION
\newcount\SUBSECTION
\newcount\FNUMBER

\PART=0
\CHAPTER=0
\SECTION=0
\SUBSECTION=0	
\FNUMBER=0

\def\LastPart{\ModeUndef}
\def\LastChapter{\ModeUndef}
\def\LastSection{\ModeUndef}
\def\LastSubSection{\ModeUndef}
\def\LastClaim{\ModeUndef}
\def\Last{\ModeUndef}

\newdimen\TOBOTTOM
\newdimen\LIMIT

\def\Ensure#1{\ \par\ \immediate\LIMIT=#1\immediate\TOBOTTOM=\the\pagegoal\advance\TOBOTTOM by -\pagetotal%
\ifdim\TOBOTTOM<\LIMIT\newpage \else%
\vskip-\parskip\vskip-\parskip\vskip-\baselineskip\fi}

\def\PartLabel{\the\PART}
\def\NewPart#1{\global\advance\PART by 1%
         \bs\ni{\PartStyle  Part \PartLabel:}
         \bs\ni{\PartStyle #1}\newpage%
         \CHAPTER=0\SECTION=0\SUBSECTION=0\FNUMBER=0%
         \gdef\Left{#1}%
         \global\edef\Last{\PartLabel}%
         \global\edef\LastPart{\PartLabel}%
         \global\edef\LastChapter{\ModeUndef}%
         \global\edef\LastSection{\ModeUndef}%
         \global\edef\LastSubSection{\ModeUndef}%
         \global\edef\LastClaim{\ModeUndef}}
\def\ChapterLabel{\the\CHAPTER}
\def\NewChapter#1{\global\advance\CHAPTER by 1%
         \bs\ni{\ChapterStyle  Chapter \ChapterLabel: #1}\ms%
         \SECTION=0\SUBSECTION=0\FNUMBER=0%
         \gdef\Left{#1}%
         \global\edef\Last{\ChapterLabel}%
         \global\edef\LastChapter{\ChapterLabel}%
         \global\edef\LastSection{\ModeUndef}%
         \global\edef\LastSubSection{\ModeUndef}%
         \global\edef\LastClaim{\ModeUndef}}
\def\SectionEnsure{3cm}
\def\NewSection#1{\Ensure{\SectionEnsure}\gdef\SectionLabel{\the\SECTION}\global\advance\SECTION by 1%
         \ms\ni{\SectionStyle  \SectionLabel.\ #1}\ss%
         \SUBSECTION=0\FNUMBER=0%
         \gdef\Left{#1}%
         \global\edef\Last{\SectionLabel}%
         \global\edef\LastSection{\SectionLabel}%
         \global\edef\LastSubSection{\ModeUndef}%
         \global\edef\LastClaim{\ModeUndef}}
\def\NewAppendix#1#2{\Ensure{\SectionEnsure}\gdef\SectionLabel{#1}\global\advance\SECTION by 1%
         \bs\ni{\SectionStyle  Appendix \SectionLabel.\ #2}\ss%
         \SUBSECTION=0\FNUMBER=0%
         \gdef\Left{#2}%
         \global\edef\Last{\SectionLabel}%
         \global\edef\LastSection{\SectionLabel}%
         \global\edef\LastSubSection{\ModeUndef}%
         \global\edef\LastClaim{\ModeUndef}}
\def\Acknowledgements{\Ensure{\SectionEnsure}\gdef\SectionLabel{}%
         \ms\ni{\SectionStyle  Acknowledgments}\ss%
         \SECTION=0\SUBSECTION=0\FNUMBER=0%
         \gdef\Left{}%
         \global\edef\Last{\ModeUndef}%
         \global\edef\LastSection{\ModeUndef}%
         \global\edef\LastSubSection{\ModeUndef}%
         \global\edef\LastClaim{\ModeUndef}}
\def\SubSectionEnsure{2cm}
\def\SubSectionLabel{\ifnum\SECTION>0 \the\SECTION.\fi\the\SUBSECTION}
\def\NewSubSection#1{\Ensure{\SubSectionEnsure}\global\advance\SUBSECTION by 1%
         \ms\ni{\SubSectionStyle #1}\ss%
         \global\edef\Last{\SubSectionLabel}%
         \global\edef\LastSubSection{\SubSectionLabel}}
\def\SetNumberingModeN{\def\ClaimLabel{(\the\FNUMBER)}}
\def\SetNumberingModeSN{\def\ClaimLabel{(\ifnum\SECTION>0 \SectionLabel.\fi%
      \the\FNUMBER)}}
\def\SetNumberingModeCSN{\def\ClaimLabel{(\ifnum\CHAPTER>0 \the\CHAPTER.\fi%
      \ifnum\SECTION>0 \SectionLabel.\fi%
      \the\FNUMBER)}}

\def\NewClaim{\global\advance\FNUMBER by 1%
    \ClaimLabel%
    \global\edef\LastClaim{\ClaimLabel}%
    \global\edef\Last{\ClaimLabel}}

\def\HideLabels{\xdef\ShowLabelsMode{\ModeNo}}
\HideLabels

\def\fn{\eqno{\NewClaim}} 
\def\fl#1{%
\ifx\ShowLabelsMode\ModeYes%
 \eqno{{\buildrel{\hbox{\AbstractStyle[#1]}}\over{\hfill\NewClaim}}}%
\else%
 \eqno{\NewClaim}%
\fi%
\dbLabelInsert{#1}{\ClaimLabel}}
\def\fprel#1{\global\advance\FNUMBER by 1\StoreLabel{#1}{\ClaimLabel}%
\ifx\ShowLabelsMode\ModeYes%
\eqno{{\buildrel{\hbox{\AbstractStyle[#1]}}\over{\hfill.\itag}}}%
\else%
 \eqno{\itag}%
\fi%
}

\def\cl#1{\global\advance\FNUMBER by 1\dbLabelInsert{#1}{\ClaimLabel}%
\ifx\ShowLabelsMode\ModeYes%
${\buildrel{\hbox{\AbstractStyle[#1]}}\over{\hfill\ClaimLabel}}$%
\else%
  $\ClaimLabel$%
\fi%
}
\def\cprel#1{\global\advance\FNUMBER by 1\StoreLabel{#1}{\ClaimLabel}%
\ifx\ShowLabelsMode\ModeYes%
${\buildrel{\hbox{\AbstractStyle[#1]}}\over{\hfill.\itag}}$%
\else%
  $\itag$%
\fi%
}

\def\Note{\ms\leftskip 3cm\rightskip 1.5cm\AbstractStyle}
\def\endNote{\par\leftskip 2cm\rightskip 0cm\NormalStyle\ss}


\parindent=7pt
\leftskip=2cm
\newcount\SideIndent
\newcount\SideIndentTemp
\SideIndent=0
\newdimen\SectionIndent
\SectionIndent=-8pt

\def\sidebar{\vrule height15pt width.2pt }
\def\endcorner{\hbox{\hbox{\vrule height6pt width.2pt}\vbox to6pt{\vfill\hbox
to4pt{\leaders\hrule height0.2pt\hfill}}}}
\def\begincorner{\hbox{\hbox{\vrule height6pt width.2pt}\vbox to6pt{\hbox
to4pt{\leaders\hrule height0.2pt\hfill}}}}
\def\endbegincorner{\hbox{\vbox to15pt{\endcorner\vskip-6pt\begincorner\vfill}}}
\def\SideShow{\SideIndentTemp=\SideIndent \ifnum \SideIndentTemp>0 
\loop\sidebar\hskip 2pt \advance\SideIndentTemp by-1\ifnum \SideIndentTemp>1 \repeat\fi}

\def\BeginSection{{\vbadness 100000 \par\ni\hskip\SectionIndent%
\SideShow\vbox to 15pt{\vfill\begincorner}}\global\advance\SideIndent by1\vskip-10pt}

\def\EndSection{{\vbadness 100000 \par\ni\global\advance\SideIndent by-1%
\hskip\SectionIndent\SideShow\vbox to15pt{\endcorner\vfill}\vskip-10pt}}

\def\EndBeginSection{{\vbadness 100000\par\ni%
\global\advance\SideIndent by-1\hskip\SectionIndent\SideShow
\vbox to15pt{\vfill\endbegincorner}}%
\global\advance\SideIndent by1\vskip-10pt}

\def\ShowBeginCorners#1{%
\SideIndentTemp =#1 \advance\SideIndentTemp by-1%
\ifnum \SideIndentTemp>0 %
\vskip-15truept\hbox{\kern 2truept\vbox{\hbox{\begincorner}%
\ShowBeginCorners{\SideIndentTemp}\vskip-3truept}}%
\fi%
}

\def\ShowEndCorners#1{%
\SideIndentTemp =#1 \advance\SideIndentTemp by-1%
\ifnum \SideIndentTemp>0 %
\vskip-15truept\hbox{\kern 2truept\vbox{\hbox{\endcorner}%
\ShowEndCorners{\SideIndentTemp}\vskip 2truept}}%
\fi%
}

\def\BeginSections#1{{\vbadness 100000 \par\ni\hskip\SectionIndent%
\SideShow\vbox to 15pt{\vfill\ShowBeginCorners{#1}}}\global\advance\SideIndent by#1\vskip-10pt}

\def\EndSections#1{{\vbadness 100000 \par\ni\global\advance\SideIndent by-#1%
\hskip\SectionIndent\SideShow\vbox to15pt{\vskip15pt\ShowEndCorners{#1}\vfill}\vskip-10pt}}

\def\EndBeginSections#1#2{{\vbadness 100000\par\ni%
\global\advance\SideIndent by-#1%
\hbox{\hskip\SectionIndent\SideShow\kern-2pt%
\vbox to15pt{\vskip15pt\ShowEndCorners{#1}\vskip4pt\ShowBeginCorners{#2}}}}%
\global\advance\SideIndent by#2\vskip-10pt}




%
%


\def\al{\alpha}
\def\be{\beta}
\def\de{\delta}
\def\ga{\gamma}

\def\ep{\epsilon}

\def\la{\lambda}

\def\om{\omega}
\def\si{\sigma}
\def\vp{\varphi}

\def\De{\Delta}
\def\Ga{\Gamma}



 \def\gotg{{\hbox{\gothic g}}}
 


 \def\R{{\hbox{\Bbb R}}}

 \def\R{{\hbox{\Bbb R}}}


\def\det{{\hbox{det}}}

\def\ip{\hbox to4pt{\leaders\hrule height0.3pt\hfill}\vbox to8pt{\leaders\vrule width0.3pt\vfill}\kern 2pt}
 
\def\del{\partial}
\def\na{\nabla}

\def\arr{\rightarrow}

\def\then{\Rightarrow}

%
%

\def\DEFINITION{\ClaimStyle\ni{\bf Definition: }}

\def\ENDDEFINITION{\NormalStyle}

\def\cases#1{\left\{\eqalign{#1}\right.}
\NormalStyle
\SetNumberingModeSN
\PreventDoubleOn

\long\def\title#1{\centerline{\TitleStyle\ni#1}}

\long\def\author#1{\ms\centerline{\AuthorStyle by {\it #1}}}

\def\abstract{\ms\leftskip 3cm\rightskip .5cm\AbstractStyle{\bf \ni Abstract:}\ }
\def\endabstract{\par\leftskip 2cm\rightskip 0cm\NormalStyle\ss}

\SetNumberingModeSN

\def\N{{\hbox{\Bbb N}}}

\def\nab#1{{\buildrel #1\over \na}}
\def\frac[#1/#2]{\hbox{$#1\over#2$}}
\def\Frac[#1/#2]{{#1\over#2}}
\def\({\left(}
\def\){\right)}
\def\[{\left[}
\def\]{\right]}
\def\^#1{{}^{#1}_{\>\cdot}}
\def\_#1{{}_{#1}^{\>\cdot}}
\def\Label=#1{{\buildrel {\hbox{\fiveSerif \ShowLabel{#1}}}\over =}}
\def\<{\kern -1pt}


\def\ExpandAllCNotes{\long\def\CNote##1{%
\BeginSection
	\Note%
 		##1%
	\endNote%
\EndSection%
}}
\ExpandAllCNotes
%
%
%
%


\def\red#1{\textcolor{red}{#1}}
\def\blue#1{\textcolor{blue}{#1}}
\def\green#1{\textcolor{green}{#1}}

\def\frame#1{\vbox{\hrule\hbox{\vrule\vbox{\kern2pt\hbox{\kern2pt#1\kern2pt}\kern2pt}\vrule}\hrule\kern-4pt}}

\def\Box to #1#2#3{\frame{\vtop{\hbox to #1{\hfill #2 \hfill}\hbox to #1{\hfill #3 \hfill}}}}


\bib{EPS}{J.Ehlers, F.A.E.Pirani, A.Schild, 
{\it The Geometry of Free Fall and Light Propagation},
in General Relativity, ed. L.OÕRaifeartaigh (Clarendon, Oxford, 1972). 
}

\bib{Perlick}{V. Perlick, 
{\it Characterization of standard clocks by means of light rays and freely falling particles}
General Relativity and Gravitation,  {\bf 19}(11) (1987) 1059-1073}

\bib{BiMetricTheories}{Komar, BiMetricTheories 
}

\bib{NoGo}{E. Barausse, T.P. Sotiriou, J.C. Miller,
{\it A no-go theorem for polytropic spheres in Palatini $f(R)$ gravity}, 
Class. Quant. Grav. {\bf 25} (2008) 062001; gr-qc/0703132
}

\bib{Faraoni}{T.P. Sotiriou, V. Faraoni,
{\it  $f (R)$  theories of gravity},
(2008); arXiv: 0805.1726v2
}

\bib{NoGo2}{G.J. Olmo,
{\it  Re-examination of polytropic spheres in Palatini $f(R)$ gravity},
Phys.Rev. D {\bf 78} (2008) 104026; gr-qc/0703132
}

\bib{Capozziello}{S. Capozziello, M. De Laurentis, V. Faraoni
{\it A bird's eye view of $f(R)$-gravity}
(2009); arXiv:0909.4672 
}

\bib{FETG2}{L. Fatibene,  M. Ferraris, M. Francaviglia, S. Mercadante,
{\it Further Extended Theories of Gravitation: Part II};
arXiv:0911.2842
}

\bib{Magnano}{G. Magnano, L.M. Sokolowski, 
{\it On Physical Equivalence between Nonlinear Gravity Theories}
Phys.Rev. D50 (1994) 5039-5059; gr-qc/9312008
}

\bib{S1}{T.P. Sotiriou, S. Liberati,
{\it Metric-affine f(R) theories of gravity},
Annals Phys. 322 (2007) 935-966; gr-qc/0604006
}

\bib{S2}{T.P. Sotiriou,
{\it $f(R)$ gravity, torsion and non-metricity},
Class. Quant. Grav. 26 (2009) 152001; gr-qc/0904.2774}

\bib{S3}{T.P. Sotiriou,
{\it Modified Actions for Gravity: Theory and Phenomenology},
Ph.D. Thesis; gr-qc/0710.4438}

\bib{C1}{S. Capozziello, M. Francaviglia,
{\it Extended Theories of Gravity and their Cosmological and Astrophysical Applications},
Journal of General Relativity and Gravitation 40 (2-3), (2008) 357-420.}

\bib{C2}{S. Capozziello, M.F. De Laurentis, M. Francaviglia, S. Mercadante,
{\it From Dark Energy and Dark Matter to Dark Metric},
Foundations of Physics 39 (2009) 1161-1176
gr-qc/0805.3642v4}

\bib{C3}{S. Capozziello, M. De Laurentis, M. Francaviglia, S. Mercadante,
{\it First Order Extended Gravity and the Dark Side of the Universe: the General Theory}
Proceedings of the Conference ``Univers Invisibile'', Paris June 29 Ð July 3, 2009 
- to appear in 2010}

\bib{C4}{S. Capozziello, M. De Laurentis, M. Francaviglia, S. Mercadante,
{\it First Order Extended Gravity and the Dark Side of the Universe Ð II: Matching Observational Data},
Proceedings of the Conference ``Univers Invisibile'', Paris June 29 Ð July 3, 2009 
Ð to appear in 2010}



\def\ubal{\underline{\al}\kern1pt}
\def\obal{\overline{\al}\kern1pt}

\def\ubR{\underline{R}\kern1pt}
\def\obR{\overline{R}\kern1pt}
\def\ubom{\underline{\om}\kern1pt}
\def\obxi{\overline{\xi}\kern1pt}
\def\ubu{\underline{u}\kern1pt}
\def\ube{\underline{e}\kern1pt}
\def\obe{\overline{e}\kern1pt}

\def\AppA{A}

\NormalStyle
\edef\PreModeStatus{\ModeYes}
	\immediate\openin\AuxFile=PreLabels.def
	\ifeof \AuxFile
	\else
 		\immediate\closeout\AuxFile
  		\input PreLabels.def
 	 \fi
	 \immediate\openout\AuxFile=PreLabels.def

\title{Further Extended Theories of Gravitation: Part I\footnote{$^\ast$}{{\AbstractStyle
	This paper is published despite the effects of the Italian law 133/08 ({\tt http://groups.google.it/group/scienceaction}). 
        This law drastically reduces public funds to public Italian universities, which is particularly dangerous for free scientific research, 
        and it will prevent young researchers from getting a position, either temporary or tenured, in Italy.
        The authors are protesting against this law to obtain its cancellation.\goodbreak}}}

\author{M.Di Mauro, L.Fatibene,  M.Ferraris, M.Francaviglia}


\abstract
We shall here propose a class of relativistic theories of gravitation, based on a foundational paper of Ehlers Pirani and Schild (EPS).
All {\it ``extended theories of gravitation''} (also known as $f(R)$ theories) in Palatini formalism are shown to belong to this class. 
In a forthcoming paper we shall show that this class of theories contains other more general examples.

EPS framework helps in the interpretation and solution of these models that however have exotic behaviours even compared to $f(R)$ theories.  
\endabstract

\NewSection{Introduction}

Once one starts to generalize GR in order to include new observational data it is not clear whether there is a natural border that physically reasonable models of gravitation should not cross.
Ehlers, Pirani and Schild (EPS) proposed in 1972
some axioms to deduce GR from observational structures; see  \ref{EPS}. 
Their subtle analysis was based on physical and mathematical axioms and turned into the understanding that gravity and causality require on spacetime both an affine and a metric structure {\it a priori} independent but related by suitable compatibility requirements.
The program was never completed in a fully satisfactory way,
in the sense that they could not finally prove that one needs to use as a connection  the Levi-Civita connection of the metric structure on spacetime.

However, EPS framework turns out to be a natural framework for the interpretation of Palatini extended theories of gravitation in general, 
and in particular for $f(R)$ models. These models have been recently used with the hope that they could account for effects attributed to dark matter and energy; \ref{Capozziello}, \ref{Faraoni}, \ref{S1}, \ref{S3}, \ref{C3}, \ref{C4} and references quoted therein.
A number of criticisms have been variously raised on Palatini framework; see e.g.~\ref{Faraoni} , \ref{NoGo} and references quoted therein. However, in view of EPS framework it appears as the most natural setting for gravitational theories, at least from a foundational viewpoint; see also \ref{C1}, \ref{C2}.

This paper is divided into two parts. In this first part we shall review the EPS, define the class of {\it further extended theories of gravitation} (FETG)
and show that $f(R)$ theories are particular cases of FETG.

In the second part we shall consider some examples of FETG other than standard $f(R)$ theories.
We shall show that FETG allow projective structures (i.e., connections)  that are not necessarily metric. Moreover, a specific example will be considered 
that reproduces the behaviour of purely metric $f(R)$ theories starting from a Palatini framework. This is particularly interesting since it overcomes some of the recent criticisms against Palatini framework based on problems in modelling polytropic stars; see \ref{NoGo}.

Hereafter, spacetime $M$ is assumed to be a 
 $4$-dimensional, connected, paracompact manifold which allows global Lorentzian metrics.

\NewSection{EPS Axioms and Compatibility}

Ehlers, Pirani and Schild (EPS) proposed in the seventies an axiomatic construction of General Theory of Relativity (see \ref{EPS}). 
They showed that spacetime geometry can be obtained from few assumptions about two observational quantities: the worldlines of {\it light rays}  and free falling  {\it mass particles}.
Light rays and particles are to be understood in a classical sense: light rays are {\it ``small wave packets''} and particles are material balls 
with negligible extension. 

In order to introduce a differential structure on $M$, EPS defined the {\it radar coordinates}:
given an event $e\in M$ and two particles (closed enough to $e$) one can consider echoes from the particles on the event $e$.
Radar coordinates are in particular defined by the following map:
$$
\Phi_{pp'}: e \mapsto (u,v,u',v')
\fn$$ 
where $(u,v,u',v')$ are the leaving and arriving parameters of echoes on the two particles.

Then EPS introduced a function $g$ (see Axiom $L_1$ in \ref{EPS}) considering a particle $P$ passing through an event $e$ and light cone $v_e$ emitted from an event $p$ external to $P$. The function $g$ is defined by $g=t(e1)t(e2)$, where $t(e1)$ and $t(e2)$ are parameters on $P$ relative to the encounter of $v_e$ with the particle $P$. 

The Hessian matrix (at $e$) of this map $g$ defines a tensor $g_{\mu\nu}$ with the following  properties:
$$
         g_{\mu\nu}T^\mu T^\nu=0
         \qquad\qquad
         g_{\mu\nu}L^\mu L^\nu=2
\fl{Uno}$$
for any vector $T$ tangent to light rays passing through $e$ and any  vector $L$ tangent to particles passing through $e$.
Unfortunately, the tensor $g_{\mu\nu}$ depends on the parametrization fixed along the particles which are unphysical;
they in fact correspond to fix a conventional clock (see \ref{Perlick}).
One can show that physically one can single out a class of tensors which are defined modulo a conformal factor
(which in fact does not affect light cones defined by $g$).

One can easily check that EPS axioms are coherent. 
For example one can start with $M=\R^4$ and two families of straight lines and show that
in Cartesian coordinates the tensor $g_{\mu\nu}$ turns out to be
$$
          g_{\mu\nu}=\frac[2/\|u\|^2]\eta_{\mu\nu}
\fn$$
where   $(\del_0 , u)$ are the $4$-velocities of the particles involved in the definition of radar coordinates and 
for $\eta_{\mu\nu}$ is the standard Minkowski metric. 
So we have found that the tensor $g_{\mu\nu}$ verifies \ShowLabel{Uno}. Moreover $\det(g_{\mu\nu})\neq{0}$ so it defines in fact a spacetime metric
which exhibits $M$ as  equivalent to Minkowski spacetime.

In order to introduce an  object  that is invariant with respect to particles reparametrizations (i.e.~conformal transformations on $g$)
EPS considered the {\it conformal metric} 
$$
          \gotg_{\mu\nu}=\Frac[g_{\mu\nu}/{}^4\<\<\sqrt{|\det g |}]
\fn$$
which is a tensor density of weight $\frac[1/2]$.

\Note
The quantity $\gotg_{\mu\nu}$ is invariant for conformal transformations $g'_{\mu\nu}=\vp \cdot g_{\mu\nu}$; in fact
$$
\gotg'_{\mu\nu}=  \Frac[g'_{\mu\nu}/{}^4\<\sqrt{g'}]= \Frac[\vp \cdot g_{\mu\nu}/{}^4\<\sqrt{\vp^4 \cdot g}]
=\Frac[g_{\mu\nu}/{}^4\<\sqrt{g}]=\gotg_{\mu\nu}
\fl{ConformalMetricTransf}$$
\endNote

For simplicity let us set $ {}^4\<\<\sqrt{g}:={}^{4}\<\<\sqrt{|\det{g}|} $. 
The {\it conformal metric}  $\gotg_{\mu\nu}$ defines a {\it conformal structure} on $M$.
The equation $\gotg_{\mu\nu}T^\mu T^\nu=0$ represents light cones emitted from (or received at) an event $e$ so it allows to distinguish among timelike, spacelike and lightlike vectors through $e$.
The conformal structure is equivalent to assigning the light cones at each point.
Hereafter in this Section, spactime indices will be lowered and raised by the conformal metric $\gotg$.

Then EPS axiom $P_2$ relates to an infinitesimal version of the law of inertia in {\it projective coordinates} $\bar x^\al$, namely
$$
\Frac[d^2\bar{x}^{\al}/d\bar{u}^2]=0
$$
Changing parameters and coordinates the previous equation becomes:
$$
          \ddot{x}^\al+\Pi^\al_{\mu\nu}\,\dot{x}^\mu\dot{x}^\nu=\lambda\dot{x}^\al
 \fl{GeodesicsEquation}$$
The coefficients $\Pi^\al_{\mu\nu}$ can be choosen without loss of generality to obey the properties $\Pi^\al_{[\mu\nu]}=0$ and $\Pi^\al_{\mu\al}=0$; they are called {\it projective coefficients}. 
On the other hand one can fix parametrization so to have $\la=0$.

They do not directly refer to a connection because in view of the traceless condition they transform differently
according to the following transformation rules:
$$
          \Pi'^\al_{\be\mu}=J^\al_\la\(\Pi^\la_{\rho\si} \bar{J}^\rho_\be \bar J^\si_\mu+\bar{J}^\la_{\be\mu}\) 
          +\frac[2/5] \de^\al_{(\be}\de^\ep_{\mu)} \bar{J}^\rho_{\ep}\del_\rho \ln J
\fl{PiTL}$$
The projective coefficients define thence a geometric object which is called a {\it projective connection} and a {\it projective structure} on $M$.
 Any curve that satisfies equation \ShowLabel{GeodesicsEquation} is said {\it geodesic} (trajectory). In particular,  particles follow geodesics.

Once the conformal and the projective structure of spacetime have been introduced,
EPS proposed a compatibility axiom (see \ref{EPS} Axiom C): 

\ss
{\leftskip 3cm\rightskip .5cm\ni\sl 
a solution of equations \ShowLabel{GeodesicsEquation} is a particle iff 
its initial $4$-velocity at $e$ is contained in the interior of the light cone $\nu_e$.\par}
\ms

This axiom allows to express an analytical relation  between the conformal and the projective structure. 
They introduced $\{\gotg\}^\al_{\mu\nu}$, the ``Christoffel symbols'' of $\gotg_{\mu\nu}$:
$$
          \{\gotg\}^\al_{\mu\nu}:=\{g\}^\al_{\mu\nu}
           +\frac[1/8] \(g^{\ga\al}g_{\mu\nu}
           -2\de^\al_{(\mu} \de^\ga_{\nu)} \) \del_{\ga}\ln g
\fn$$
where $g$ is any representative of the conformal class identified by $\gotg$.
The coefficients $\{\gotg\}^\al_{\mu\nu}$ transform as:
$$
           \{\gotg'\}^\al_{\be\mu}=J^\al_\la( \{\gotg\}^\la_{\rho\si} \bar{J}^\rho_\be \bar{J}^\si_\mu+\bar{J}^\la_{\be\mu})
           -\frac[1/4] J^\al_\la \( \gotg^{\la\ep } \gotg_{\rho\si} -2\de^\la_{(\rho}\de^\ep_{\si)}  \)  \bar J^\rho_\be \bar J^\si_\mu \del_\ep \ln J 
\fl{KTL}$$

The difference $\De^\al_{\be\mu}=\Pi^\al_{\be\mu}-\{\gotg\}^\al_{\be\mu}$ has the following two properties $\De^\al_{[\be\mu]}=0$ and $\De^\al_{\be\al}=0$.
Then EPS obtained  the relation for $\De_{\al\be\mu}:= \gotg_{	\al\la} \De^\la_{\be\mu}$:
$$
           \De_{\al\be\ga}=\De_{(\al\be\ga)}+\frac[1/2](p_\al \gotg_{\be\ga}-\gotg_{\al(\be}p_{\ga)})+L_{\al(\be\ga)}
\fl{EPS1}$$
where we set $L_{\al\be\ga}=\frac[4/3]\Delta_{[\al\be]\ga}-p_{[\al} \gotg_{\be]\ga}$ and $p_\al=\frac[8/9]\Delta_{[\al\la]}{}^{\la}$. 
(In \ref{EPS} the definition of $p_\al$ was given with a different sign probably due to a typo. 
With our definition the quantity $L_{\al\be\ga}$ has the following properties
$L^\al{}_{\be\al}=0$, $L_{[\al\be\ga]}=0$ e $L_{(\al\be)\ga}=0$ 
while with the original EPS definition  the first one does not hold true; see Appendix \AppA).

Equation \ShowLabel{EPS1} can be recasted into the following form
$$
           \De^\al_{\be\mu}=L^\al_{(\be\mu)}+5 \tilde q^\al \gotg_{\be\mu}-2\de^\al_{(\be}\tilde q_{\mu)}
\fn$$
by setting $\tilde q^\al=\frac[1/18]\De^\al_{\be\mu} \gotg^{\be\mu}$. 

Now consider a sequence $\{P_n: n\in \N\}$ of particles passing through an event $e$. For Axiom C all $P_n$ are internal to the light cone $v_e$. $P_n$ are particles (hence projective geodesics). Let us consider a sequence determined by initial velocities that tend to a light velocity $c^\mu$.
Let $P$ be the light ray with initial velocity $c^\mu$.
Since $P_n \arr P$, the $4$-velocity of particles can be as near as one wishes to $c^\mu$; accordingly, $P$ is a projective geodesic {\it and} a conformal geodesics.
Hence Axiom C can be equivalently expressed in this way: 

\ss
{\leftskip 3cm\rightskip .5cm\ni\sl 
null projective-geodesic are identical to null conformal-geodesics.\par}
 \ms

Conformal geodesics are obtained as solution of the following equation:
$$
           \ddot{x}^\al+\{\gotg\}^\al_{\mu\nu}\,\dot{x}^\mu\dot{x}^\nu=\nu\,\dot{x}^\al
\fn$$
EPS were at this point able to give a formal expression to the compatibility condition as
$$
           \De^\al_{\mu\nu}=5\tilde q^\al \gotg_{\mu\nu}-2\delta^\al_{(\mu}\tilde q_{\nu)}
\fl{EPS4}$$
Another equivalent form of Axiom C is: 

\ss
{\leftskip 3cm\rightskip .5cm\ni\sl 
 the set of particles is identical with the set of conformal-timelike  projective-geodesics.\par}
 \ms

Finally, EPS introduced a connection $\Ga^\al_{\be\mu}$ given in terms of the projective and conformal structures
$$
           \Gamma^\al_{\be\mu}:=\{\gotg\}^\al_{\be\mu}+ \(\gotg^{\al\ep}\gotg_{\be\mu}-2\de^\al_{(\be}\de^\ep_{\mu)}\) q_\ep
\fl{EPSConnection}$$
where we set $q_\ep:= 5\tilde q_\ep$.

\Note
This can be introduced as a generic linear combination 
$$
 \Gamma^\al_{\be\mu}=a\,\{\gotg\}^\al_{\be\mu}+b\tilde q^\al \gotg_{\be\mu}-2c\,\delta^\al_{(\be} \tilde q_{\mu)}
\fn$$
 and then fixing the coefficients so that $\Ga$ transforms
as a connection.
In view of  \ShowLabel{PiTL} and \ShowLabel{KTL} one can check that $\Ga$ transfoms in fact as a connection
$$
 \eqalign{
   &   \Gamma'^\al_{\be\mu}=a\,\{\gotg\}'^\al_{\be\mu}+b \tilde q'^\al \gotg'_{\be\mu}-2c\,\delta^\al_{(\be} \tilde q'_{\mu)}= \cr
    &=  J^\al_\la \(  \Gamma'^\la_{\rho\si} \bar J^\rho_\be \bar J^\si_\mu + a \bar J^\la_{\be\mu}\)
    +J^\al_\la \(  \(\frac[b/20]-\frac[a/4]\) \gotg^{\la\ep}\gotg_{\rho\si}
    	+\(\frac[a/2] -\frac[c/10]\) \de^\la_{(\rho}\de^\ep_{\si)}\) \bar J^\rho_\be \bar J^\si_\mu \del_\ep \ln J
}
\fn$$
This transforms as a connection iff we set $a=1$ and  $b=c=5$. Accordingly we have \ShowLabel{EPSConnection}.
\endNote

\

\NewSection{Non-metric EPS Connections}

EPS showed that the most general connection $\Ga^\al_{\be\mu}$ compatible (in EPS sense) with a metric structure $g_{\mu\nu}$ is 
in the form \ShowLabel{EPSConnection}. By tracing the EPS connection we obtain
$$
\eqalign{
\Ga^\al_{\al\mu}=& \{\gotg\}^\al_{\al\mu} +  \( \gotg^{\al\ep}\gotg_{\al\mu} - \de^\al_{\al} \de^\ep_{\mu}- \de^\al_{\mu} \de^\ep_{\al} \)q_{\ep}
= \{\gotg\}^\al_{\al\mu} +  \( \red{\de^\ep_\mu}  - 4\de^\ep_{\mu}- \red{\de^\ep_{\mu} } \)q_{\ep}=\cr
=& \{\gotg\}^\al_{\al\mu}  - 4q_{\mu}
}
\fn$$

\Note
If $h_{\mu\nu}=\vp \cdot g_{\mu\nu}$ we have moreover
$$
\eqalign{
\{ h \}^\al_{\be\mu}=&\frac[1/2] h^{\al\la}\(-\del_{\la} h_{\be\mu}+\del_{\be} h_{\mu\la}+\del_{\mu} h_{\la\be} \)=
\{g \}^\al_{\be\mu} + \frac[1/2\vp ] g^{\al\la}\(-\del_{\la} \vp g_{\be\mu}+\del_{\be} \vp g_{\mu\la}+\del_{\mu} \vp g_{\la\be} \)=\cr
=&\{g \}^\al_{\be\mu} + \frac[1/2 ] \(-\del_{\la} \ln\vp g^{\al\la}g_{\be\mu}+\del_{\be} \ln\vp \de^\al_\mu +\del_{\mu} \ln\vp \de^\al_\be \)=\cr
=&\{g \}^\al_{\be\mu} - \frac[1/2 ] \( g^{\al\la}g_{\be\mu}-2 \de^\la_{(\be}\de^\al_{\mu)}  \)\del_{\la} \ln\vp\cr
}
\fl{LCConformalMetric}$$

The case $h_{\mu\nu}=\gotg_{\mu\nu}= \frac[g_{\mu\nu}/{}^4\<\sqrt{g}]$ is a special case for $\vp=g^{-1/4}$;
$$
\eqalign{
\{ \gotg \}^\al_{\be\mu}=&\{g \}^\al_{\be\mu} + \frac[1/8 ] \( g^{\al\la}g_{\be\mu}-2 \de^\la_{(\be}\de^\al_{\mu)}  \)\del_{\la} \ln g
}
\fl{ConformalChristoffel}$$
Notice the similarity with \ShowLabel{EPSConnection} for $q_\ep=\frac[1/8 ] \del_{\la} \ln g$.
The trace is:
$$
\eqalign{
\{ \gotg \}^\al_{\al\mu}=&\{g \}^\al_{\al\mu} + \frac[1/8 ] \( g^{\al\la}g_{\al\mu}- \de^\la_{\al}\de^\al_{\mu}
- \de^\la_{\mu}\de^\al_{\al}  \)\del_{\la} \ln g=\cr
=&\frac[1/2 ]g^{\al\ep}\(-\red{\del_{\ep} g_{\al\mu}}+\red{\del_{\al} g_{\mu\ep}}+\del_{\mu} g_{\ep\al}\) 
+ \frac[1/8 ] \( \green{\de^\la_\mu} - \green{\de^\la_{\mu}}- 4\de^\la_{\mu}  \)\del_{\la} \ln g=\cr
=&\frac[1/2g ] \del_{\mu}  g - \frac[1/2 ]  \del_{\mu} \ln g=
\frac[1/2 ] \del_{\mu} \ln g - \frac[1/2 ]  \del_{\mu} \ln g\equiv 0\cr
}
\fn$$
\endNote

Thence we have 
$$
q_{\mu}=- \frac[1/4] \Ga^\al_{\al\mu}
\fn$$ 

Moreover, using \ShowLabel{ConformalChristoffel} we have
$$
\Ga^\al_{\be\mu}= \{g \}^\al_{\be\mu}  + \( g^{\al\ep}g_{\be\mu} - 2\de^\al_{(\be} \de^\ep_{\mu)} \)
\(q_{\ep} + \frac[1/8 ]\del_{\ep} \ln g\)
\fl{EPSComp}$$
If we define a conformal metric $h_{\mu\nu}=\vp\cdot g_{\mu\nu}$ we also have
$$
\eqalign{
\Ga^\al_{\be\mu}=& \{ h \}^\al_{\be\mu}  + \( h^{\al\ep}h_{\be\mu} - 2\de^\al_{(\be} \de^\ep_{\mu)} \)
\(q_{\ep} + \frac[1/8 ]\del_{\ep} \ln g +\frac[1/2] \del_{\la} \ln\vp\)=\cr
=& \{ h \}^\al_{\be\mu}  + \( h^{\al\ep}h_{\be\mu} - 2\de^\al_{(\be} \de^\ep_{\mu)} \)
\(q_{\ep} + \frac[1/8 ]\del_{\ep} \ln h \)\cr
}
\fn$$
i.e.~the characterization of EPS-connections is conformally invariant.

As it is well--known, the functions $q_\ep$ parametrize the  connections EPS-compatible to $g_{\mu\nu}$.
When $q_\ep= \frac[1/8]\del_\ep \ln \vp$  for some function $\vp$  then we get a metric connection $\Ga^\al_{\be\mu}= \{\vp\cdot g \}^\al_{\be\mu}$ 
for a conformal factor $\vp$.
On the other hand when $q_\ep$ has no potential then the connection $\Ga^\al_{\mu\nu}$ is non-metric.

\Note
In fact if the connection $\Ga$ is the Levi-Civita connection of some metric $\ga$ then one has
$$
\Ga^\al_{\be\al}=\frac[1/2] \ga^{\al\la}\(-\red{\del_\la \ga_{\be\al}}+ \del_\be\ga_{\al\la}+ \red{\del_\al \ga_{\la\be}}\)=
\frac[1/2\ga] \ga\ga^{\al\la} \del_\be\ga_{\al\la}
=\frac[1/2\ga]  \del_\be \ga = \del_\be \ln \sqrt\ga 
\fn$$ 
On the other hand if the same connection $\Ga$ is EPS-compatible with the metric $g$ then $\Ga^\al_{\be\al}= -4q_\be$;
hence we have the relation
$$
q_\be= -\frac[1/8]\del_\be \ln \sqrt\ga
\fn$$

Then a connection $\Ga$ which is EPS-compatible with a metric $g$ is metric or, equivalently, iff
it is the Levi-Civita connection of a metric $h$ conformal to $g$ iff $q_\ep$ has a potential.
It is non-metric otherwise.
\endNote

\NewSection{EPS-Compatibility Condition}

We need a condition to express EPS-compatibility in differential form. 
We shall then look for relativitic theories in which field equations imply EPS-compatibility.

Let us start by considering the quantity $\nab{\Ga}_\mu\(\sqrt{h} h^{\al\be}\)$ when $\Ga$ is EPS-compatible with $h$
(or any metric $g$ conformal to $h$); 
we have
$$
\eqalign{
\nab{\Ga}_\mu\(\sqrt{h} h^{\al\be}\)=&  \nab{h}_\mu\(\sqrt{h} h^{\al\be}\) 
+ K^\al_{\ep\mu} \sqrt{h} h^{\ep\be}+ K^\be_{\ep\mu} \sqrt{h} h^{\al\ep} - K^\ep_{\ep\mu} \sqrt{h} h^{\al\be}=\cr
=&  -\sqrt{h} \( h_{\rho\la}  h^{\al\be}-2\de^{(\al}_\la \de^{\be)}_\rho \) h^{\ep\rho} K^{\la}_{\ep\mu} \cr}
\fl{EPSeq}$$
where we set $ K^\al_{\ep\mu}:=  \Ga^\al_{\ep\mu}- \{h\}^\al_{\ep\mu}$; if $\Ga$ is EPS-compatible with $h$ then we have
$$ 
K^\la_{\ep\mu}=   \( h^{\la\si}h_{\ep\mu} - 2\de^\la_{(\ep} \de^\si_{\mu)} \)
\(q_{\si} + \frac[1/8 ]\del_{\si} \ln h \)
\fn$$
and substituting back into \ShowLabel{EPSeq} we obtain
$$
\eqalign{
\nab{\Ga}_\mu\(\sqrt{h} h^{\al\be}\)=&  
 -\sqrt{h} \( h_{\rho\la}  h^{\al\be}-2\de^{(\al}_\la \de^{\be)}_\rho \) h^{\ep\rho} 
\( h^{\la\si}h_{\ep\mu} - 2\de^\la_{(\ep} \de^\si_{\mu)} \)
\(q_{\si} + \frac[1/8 ]\del_{\si} \ln h \)=\cr
=& -\sqrt{h}
\( h_{\rho\la}  h^{\al\be}-\de^{\al}_\la \de^{\be}_\rho-\de^{\be}_\la \de^{\al}_\rho \) 
\( h^{\la\si}\de^\rho_\mu - h^{\la\rho}  \de^\si_{\mu}- h^{\si\rho} \de^\la_{\mu}  \)
\(q_{\si} + \frac[1/8 ]\del_{\si} \ln h \)=\cr
=& -\sqrt{h} \Big(
  \red{h^{\al\be}\de^\si_\mu} -  4 h^{\al\be}  \de^\si_{\mu}- \red{h^{\al\be}  \de^\si_{\mu}}- \green{\de^{\be}_\mu h^{\al\si} }
+   h^{\al\be}  \de^\si_{\mu}+ \blue{h^{\si\be} \de^\al_{\mu}}- \blue{h^{\be\si} \de^{\al}_\mu} +\cr
&+  h^{\be\al}  \de^\si_{\mu}
+ \green{ h^{\si\al} \de^\be_{\mu}}
\Big)\(q_{\si} + \frac[1/8 ]\del_{\si} \ln h \)
=  2\sqrt{h} 
 h^{\al\be}  \(q_{\mu} + \frac[1/8 ]\del_{\mu} \ln h \) \cr}
\fn$$
Hence the necessary condition for a connection $\Ga$ to be EPS-compatible with the metric $h$ (or any other metric $g$ conformal to $h$)
is that  
$$
\nab{\Ga}_\mu\(\sqrt{h} h^{\al\be}\)=\al_\mu \sqrt{h}  h^{\al\be}  
\fl{EPSCond}$$
Viceversa, if \ShowLabel{EPSCond} holds true then we can define
$$
q_{\mu} := \frac[1/2]\al_\mu- \frac[1/8 ]\del_{\mu} \ln h
\qquad
\tilde\Ga^\al_{\be\mu}:=  \{ h \}^\al_{\be\mu}  + \( h^{\al\ep}h_{\be\mu} - 2\de^\al_{(\be} \de^\ep_{\mu)} \)
\(q_{\ep} + \frac[1/8 ]\del_{\ep} \ln h \)
\fl{CompConnection}$$

Then one has of course that also for this connection the following holds true 
$$
\nab{\tilde\Ga}_\mu\(\sqrt{h} h^{\al\be}\)=\al_\mu \sqrt{h}  h^{\al\be}  
\fl{EPSCondTilde}$$

Let us now define the tensor  $\tilde  H^\al_{\be\mu}:= \Ga^\al_{\be\mu}-\tilde \Ga^\al_{\be\mu}$; 
by subtracting \ShowLabel{EPSCond} and \ShowLabel{EPSCondTilde} we obtain
$$
\tilde H^\al_{\ep\mu}  h^{\ep\be}+ \tilde H^\be_{\ep\mu}  h^{\al\ep} - \tilde H^\ep_{\ep\mu} h^{\al\be}=0
\fl{Unique1}$$
By tracing this last equation with $h_{\al\be}$ we obtain
$$
\tilde H^\al_{\al\mu} + H^\be_{\be\mu}  - 4H^\ep_{\ep\mu}  =0
\quad\then
\tilde H^\al_{\al\mu} =0
\fn$$
and substituting back into \ShowLabel{Unique1}
$$
\tilde H^{(\al}_{\ep\mu}  h^{\be)\ep}=0
\fl{Unique2}$$

One can now choose coordinates so that at a point $x\in M$ one has $h_{\mu\nu}(x)=\eta_{\mu\nu}$;
one can prove algebraically that \ShowLabel{Unique2} implies $\tilde H^{\al}_{\ep\mu} (x)=0$.
This can be done at any point $x\in M$ and since equation \ShowLabel{Unique2} is tensorial then $\tilde H^{\al}_{\ep\mu}=0$ holds
everywhere and in any coordinate system.
Hence one has necessarily $\Ga^\al_{\be\mu}=\tilde \Ga^\al_{\be\mu}$; 
in other words the connection $\Ga^\al_{\be\mu}$ that obeys \ShowLabel{EPSCond} is necessarily in the form:
$$
\Ga^\al_{\be\mu}:=  \{ h \}^\al_{\be\mu}  +\frac[1/2] \( h^{\al\ep}h_{\be\mu} - 2\de^\al_{(\be} \de^\ep_{\mu)} \)\al_{\ep} 
\fl{CompConnection2}$$

\ss
We are hence able to define:
\ss
\DEFINITION
 a {\it Further Extended Gravitational Theory} (FEGT) as a metric-affine relativistic theory
 (i.e.~a Lagrangian theory on a triple $(M, g, \Ga)$ with $g$ a Lorentzian metric and $\Ga$ a linear connection {\it a priori} independent of $g$)
defined by an action 
$$
L=L(g, R(\Ga)) +L_m(g, \Ga, \phi)
\fn$$
such that field equations imply EPS-compatibility, i.e.~that 
$$
\nab{\Ga}_\mu\(\sqrt{h} h^{\al\be}\)=\al_\mu \sqrt{h}  h^{\al\be}  
\fn$$
for some metric $h= \vp(\phi) \cdot g$ conformal to $g$ and some $1$-form $\al=\al_\mu(g, \phi) dx^\mu$ functions of $g$ and  the matter field $\phi$ 
(possibly together with their derivatives up to some finite order).
\ENDDEFINITION
\ms

Notice that this encompasses usual $f(R)$ theories (in Palatini formulation) in which the matter Lagrangian
is usually assumed to be independent of the connection $\Ga$.
We shall hereafter show that there are in fact non-trivial FEGT, i.e.~FEGT other that the usual $f(R)$ theories.

In such a kind of FEGT the connection $\Ga$ is defined as in \ShowLabel{CompConnection2}.
If the differential form $\al:=\al_\mu dx^\mu$ is not closed then the theory is non-trivial and its connection is non-metrical
though it turns out to be EPS-compatible with the original metric structure $g$ (or equivalently to any conformal metric structure $h$).

In such theories the lightlike geodesics coincide with the metric geodesics, but the timelike geodesics
(i.e.~worldlines of matter points) are exotic. This situation has been considered in literature (see \ref{S2}) though not in connection with EPS criteria.
Of course, further investigations should be devoted to the possibility 
that such exotic dynamics can model anomalous rotation curves in galaxies and/or cosmological acceleration
(as is already proven with $f(R)$ theories) or be detected by solar system experiments.

\ 

\NewSection{A ``Trivial'' Example: $f(R)$ Theories}

We shall hereafter inestigate the class of FETG. One class of examples is well-known: in any $f(R)$ theory, in Palatini formulation one has a connection $\Ga^{\al}_{\be\mu}=\{h\}^{\al}_{\be\mu}$ which is the Levi-Civita connection of a conformal metric $h_{\mu\nu}=f'(R(\Ga, g)) \cdot g_{\mu\nu}= f'(T) \cdot g_{\mu\nu}$, where $T=T_{\mu\nu} g^{\mu\nu}$; see \ref{Magnano}. 
Thus $\Ga^{\al}_{\be\mu}$, in view of \ShowLabel{LCConformalMetric}, has the following expression:
$$
\eqalign{
   &   \Ga^{\al}_{\mu\nu}=\{f' \cdot g\}^{\al}_{\mu\nu}=
    \{g\}^\al_{\mu\nu}-\frac[1/2] \( g^{\al\ep}g_{\mu\nu}- 2\delta^\al_{(\mu}\delta^\ep_{\nu)}\)\del_{\ep}\ln f'
}
\fl{fRGa}$$
Now we calculate the trace $\Ga^\al_{\mu\al}$:
$$    
\Ga^\al_{\mu\al}=\frac[1/2]g^{\ep\la}\del_{\mu}g_{\ep\la}-\frac[1/2] \(1-5\) \del_{\mu}\ln f'
=\frac[1/2]\del_{\mu}\ln g+2  \del_{\mu}\ln f'=  \del_{\mu}\ln (f'^2 \sqrt{g})=  \del_{\mu}\ln \sqrt{h}
\fn$$
and thence $q_\ep= -\frac[1/8] \del_{\ep}\ln h$.

The connection $\Ga^{\al}_{\mu\nu}$ is in fact compatible in the sense of EPS with the conformal structure identified by $g$.

\Note
To check this  one can simply check that the identity \ShowLabel{EPSComp} 
$$
\Ga^\al_{\be\mu}= \{g \}^\al_{\be\mu}  + \( g^{\al\ep}g_{\be\mu} - 2\de^\al_{(\be} \de^\ep_{\mu)} \)
\(q_{\ep} + \frac[1/8 ]\del_{\ep} \ln g\)
\fl{EPSComp2}$$
holds true.

Hence, by comparing \ShowLabel{fRGa} and \ShowLabel{EPSComp2}, one has to check that
$$
-\red{\frac[1/2]  \del_{\ep}\ln f'}=  q_{\ep} + \frac[1/8 ]\del_{\ep} \ln g=  -\frac[1/8] \del_{\ep}\ln h + \frac[1/8 ]\del_{\ep} \ln g=
 -\red{\frac[1/2] \del_{\ep}\ln f'} -\blue{\frac[1/8] \del_{\ep}\ln g} + \blue{\frac[1/8 ]\del_{\ep} \ln g}
\fn$$
\endNote

We can also check that compatibility condition \ShowLabel{EPSCond} holds true directly; we have that
$$
\eqalign{
H^\al_{\be\mu}=&\Ga^\al_{\be\mu}- \{g \}^\al_{\be\mu} = -\frac[1/2]  \( g^{\al\ep}g_{\be\mu} - 2\de^\al_{(\be} \de^\ep_{\mu)} \) \del_{\ep}\ln f' 
}
\fn$$
Thus we have
$$
\eqalign{
    \nab{\Ga}_\al(\sqrt{g}g^{\mu\nu})=&
  \nab{g}_\al(\sqrt{g}g^{\mu\nu}) +H^\mu_{\la\al}\sqrt{g}g^{\la\nu}+H^\nu_{\la\al}\sqrt{g}g^{\mu \la}-H^\la_{\la\al}\sqrt{g}g^{\mu\nu}=\cr
  &=  \frac[\sqrt{g}/2]\Big(  \( -\red{g^{\mu\ep} \de^\nu_\al} +  \de^\ep_{\al} g^{\mu\nu}+ \blue{\de^\mu_{\al}  g^{\ep\nu}}\) 
  + \(- \blue{g^{\nu\ep}\de^\mu_\al}  +  \de^\ep_{\al}g^{\mu \nu} + \red{\de^\nu_{\al} g^{\mu \ep}}\) +\cr
 & +\( \green{\de^\ep_\al g^{\mu\nu}} - 4 \de^\ep_{\al}g^{\mu\nu} - \green{\de^\ep_\al g^{\mu\nu}} \) \Big) \del_{\ep}\ln f' =
 -   \sqrt{g}g^{\mu\nu}  \del_{\al}\ln f' =\al_\al\sqrt{g}g^{\mu\nu}  \cr
}
\fn$$
which is in fact in the expected form with $\al= -\del_{\al}\ln f'  dx^\al= d(-\ln f'  )$. The form $\al$ is exact, in view of the fact that the connection $\Ga$
is metric by construction.

\NewSection{Conclusions and Perspectives}

We here reviewed EPS framework. In particular we corrected a misprint in the original paper and we proved identity \ShowLabel{EPS1} proving also 
that it is essentially unique (modulo two inessential parameters).
We also find an equivalent, covariant characterization of EPS compatibility which will be expected to be consequence of field equations in FETG.
In this models the connection is intially independent of the spacetime metric, but it is guaranteed to be on-shell  EPS compatible with the metric.

EPS compatibility is used in two ways: first it enhances a physical interpretation in terms of observational quantities such as light rays and free falling mass particles.
On the other hand, EPS compatibility allows to determine the connection in terms of metric and  matter fields.

We finally showed that $f(R)$ are examples of FETG. We called this example as ``trivial'' FETG. In the second part (see \ref{FETG2}) we shall present 
and discuss non-trivial examples of FETG.

\NewAppendix{\AppA}{EPS Compatibility}

EPS introduced in \ref{EPS} relation \ShowLabel{EPS1} for $\De^\al_{\mu\nu}$ using $L_{\al\be\ga}=\frac[4/3]\Delta_{[\al\be]\ga}-p_{[\al}\gotg_{\be]\ga}$ and $p_\al=-\frac[8/9]\Delta_{[\al\la]}{}^{\la}$. 
Moreover they assert that:
$$
  L^\la_{\be\la}=0 \qquad L_{[\al\be\ga]}=0 \qquad L_{(\al\be)\ga}=0
\fl{A1}$$
which are used later on to deduce equation \ShowLabel{EPS4} which is important for deducing the compatibility condition \ShowLabel{EPSConnection}.
However, considering that:
$$
  L_{\al\be\ga}=\frac[2/3](\De_{\al\be\ga}-\De_{\be\al\ga})+\frac[2/9](\Delta^{\la}_{\al\la}\gotg_{\be\ga}-\Delta^{\la}_{\be\la}\gotg_{\al\ga})
\fn$$
we find that the first relation of \ShowLabel{A1} is not verified:
$$
 L^\la_{\be\la}=\frac[2/3](\De^{\la}_{\be\la}-\De_{\be\la}{}^{\la})+\frac[2/9](\Delta^{\la}{}_\ep{}^\ep\gotg_{\be\la}-\De_{\be\ep}{}^{\ep}\delta^\la_\la)=-\frac[4/3]\Delta_{\be\ep}{}^{\ep}\not\equiv0
\fn$$

Thus in order to obtain the correct definitions of the objects involved we started from general linear combinations 
$$
       \hat L_{\al\be\ga}=2b\De_{[\al\be]\ga}+2c \hat p_{[\al}\gotg_{\be]\ga}
\fn$$
and we set $\hat p_\al=2a\De_{[\al\la]}{}^{\la}= a \De_\al$ and $\De_\al=\De_{\al\mu\nu} \gotg^{\mu\nu}$. 
Then we determined the unknown coefficients $(a, b, c)$ so that 
the required properties \ShowLabel{A1} hold true.

The second and third property in \ShowLabel{A1}  are trivially satisfied; for the first one we have:
$$
\eqalign{
    &  \hat L^\la_{\be\la}=b(\De^{\la}{}_{\be\la}-\De_{\be})+ac(\De^\la{}\gotg_{\be\la}-\De_\be\de^\la_\la)=\cr
    &  =-(3ac+b)\De_{\be}\equiv 0 \quad\then b=-3ac
}
\fn$$

So the general expression of $\hat L_{\al\be\ga}$ and $\hat p_\al$ that satisfies  \ShowLabel{A1}  is:
$$
        \hat L_{\al\be\ga}=-2ac \( 3\Delta_{[\la\be]\ga}- \De_{[\al}\gotg_{\be]\ga} \)  
        \qquad
        \hat p_\al=a\De_{\al}
\fl{hatQuantities}$$
These objects are defined in order to be able to prove the identity \ShowLabel{EPS1}.
Let us consider a linear combination
$$
        \De_{\al\be	\ga}=3 A\De_{(\al\be\ga)}+B \hat p_\al\gotg_{\be\ga}+2C\gotg{g}_{\al(\be} \hat p_{\ga)}+2D L_{\al(\be\ga)}
\fn$$
and we search for the most general set of  coefficients $(A,B,C,D)$ for which this is an identity.

By using \ShowLabel{hatQuantities}  one finds:
$$
\eqalign{
  \De_{\al\be\ga}
   =& (A-6 ac D) \De_{\al\be\ga} +  (A+3ac D) \De_{\be\ga\al}+  (A+3ac D) \De_{\ga\al\be}+\cr
 &  +a(B+ 2c D) \De_\al \gotg_{\be\ga} +a(C-c D) \De_\ga \gotg_{\al\be}+ a(C-c D) \De_\be \gotg_{\ga\al}
}
\fn$$
Choosing $(\al,\nu)$ as free parameters, the general solution is
$$
\cases{
&A=\frac[1/3]\cr
&c=-\frac[1/9aD]
}
\qquad
\cases{
&C=-\frac[1/9a]\cr
&B=\frac[2/9a]\cr
}
\fn$$
and the identity is found in the form 
$$
\De_{\al\be\ga}=\De_{(\al\be\ga)}+\frac[2/9a](\hat p_\al \gotg_{\be\ga}-\gotg_{\al(\be}\hat p_{\ga)})+2D L_{\al(\be\ga)}
\fn$$
where $(a, D)$ are free parameters.
Above in the paper we chose the particular solution $a=\frac[4/9]$, $D=\frac[1/2]$ and thence
$$
\cases{
&A=\frac[1/3]\cr
&c=-\frac[1/2]
}
\qquad
\cases{
&C=-\frac[1/4]\cr
&B=\frac[1/2]\cr
}
\fn$$

\Acknowledgements

We wish to thank G.Magnano for useful discussions.
This work is partially supported by MIUR: PRIN 2005 on {\it Leggi di conservazione e termodinamica in meccanica dei continui e teorie di campo}.  
We also acknowledge the contribution of INFN (Iniziativa Specifica NA12) and the local research funds of Dipartimento di Matematica of Torino University.

\ShowBiblio

\end